\documentclass{PoS}

\usepackage{graphicx}
\usepackage{cite}
\usepackage{bm}
\usepackage{graphicx}
\usepackage{pgf}
\usepackage{amsmath}
\usepackage{xcolor}

\usepackage{tikz}
\usetikzlibrary{shapes,arrows}
\usepackage{picture}

\title{NNLO QCD corrections to dijet production at hadron colliders}

\ShortTitle{NNLO QCD corrections to dijet production at hadron colliders}

\author{\speaker{Jo\~ao Pires}, Aude Gehrmann-De Ridder\\
        Institute for Theoretical Physics, ETH Zurich, 8093 Zurich, Switzerland\\
        E-mail: \email{pires@itp.phys.ethz.ch}, \email{gehra@phys.ethz.ch}}

\author{James Currie, Thomas Gehrmann\\
        Institute for Theoretical Physics, University of Zurich, Winterthurerstrasse 190, CH-8057\\
        E-mail: \email{jcurrie@physik.uzh.ch} ,\email{thomas.gehrmann@uzh.ch}}

\author{Nigel Glover\\
        Institute for Particle Physics Phenomenology, University of Durham, South Road, Durham DH1 3LE, England\\
        E-mail: \email{e.w.n.glover@durham.ac.uk}}

\abstract{In this talk we present the calculation of next-to-next-to-leading order (NNLO) QCD corrections to dijet production and related observables at hadron colliders in the purely gluonic channel.
Results for this channel are obtained keeping all orders of $N_C$ in the colour expansion. We show that the NNLO correction significantly reduces the scale uncertainty compared to
next-to-leading order (NLO). 
We present the NNLO double-differential single jet inclusive cross section where jets are reconstructed using the anti-$k_T$ jet algorithm and show the dependence on the 
jet resolution parameter $R$. }

\FullConference{11th International Symposium on Radiative Corrections (Applications of Quantum Field Theory to Phenomenology) (RADCOR 2013),\\
		22-27 September 2013\\
		Lumley Castle Hotel, Durham, UK }

\begin{document}
\section{Introduction}
At a hadron-hadron collider the production cross section for a hard scattering process initiated by two hadrons with momenta $P_{1}$ and $P_{2}$ can be
written in the following factorised form

\begin{eqnarray}
\label{eq:cross}
\sigma(P_1,P_2)=\sum_{i,j}\int dx_1dx_2f_i(x_1,\mu_F^2)f_j(x_2,\mu_F^2)\hat{\sigma}_{ij}(\alpha_s(\mu_R),Q^2/\mu_R^2,Q^2/\mu_F^2)
\end{eqnarray}
where we require that the scale of the reaction (for example in the production of a high-$p_T$ jet) $Q^2\gg\Lambda_{\textrm{hadronic}}\sim$200~MeV. For events with such large
momentum transfer we can neglect the effects related to the binding of the partons in the incoming hadrons and the quarks and gluons behave as free particles in the collision. The hadronic
cross section factorizes into a product of non-perturbative parton distribution functions $f_i(x,\mu_F^2)$, which describe the probability of finding a parton $i$ in the incoming hadron with a longitudinal
momentum fraction $x$, with a partonic cross section $\hat{\sigma}_{ij}$ which describes the probability for the initial-state partons to interact and produce a final-state, X where the sum 
over $i,j$ runs over the parton species in the colliding hadrons.

Since the coupling is small at high energy, the partonic cross section is calculable within perturbative QCD and has a series expansion in the strong coupling constant such that we can obtain a physical prediction
by truncating the fixed-order expansion at a given order. For the dijet production process the LO cross section starts at ${\cal O}(\alpha_s^2)$ such that at NNLO we have,
\begin{eqnarray}
\sigma(P_1,P_2)&=&\sum_{i,j}\int{\rm d}x_{1}{\rm d}x_{2}f_{i}(x_{1},\mu_F^2)f_{j}(x_{2},\mu_F^2)\Bigg[\left(\frac{\alpha_s(\mu_R)}{2\pi}\right)^2\hat{\sigma}_{ij}^{LO}+\nonumber\\
&+&\left(\frac{\alpha_s(\mu_R)}{2\pi}\right)^3\hat{\sigma}_{ij}^{NLO}(\mu_{R},\mu_{F})+\left(\frac{\alpha_s(\mu_R)}{2\pi}\right)^4\hat{\sigma}_{ij}^{NNLO}(\mu_{R},\mu_{F})
+{\cal O}(\alpha_s^5)\Bigg]\;.
\label{eq:Xsec}
\end{eqnarray}

In Eqs.~(\ref{eq:cross}) and~(\ref{eq:Xsec}) we note the appearance of unphysical parameters $\mu_F,\mu_R$ related to the inclusion of higher order effects in the theory prediction. On the one hand, the factorization
scale $\mu_F$ is related to the separation of long and short distance physics as it is the scale that defines the subtraction of initial-state collinear divergences
into the non-perturbative parton distribution functions $f_i(x,\mu_F^2)$. On the other hand, the renormalization scale $\mu_R$ is introduced by ultraviolet (UV) counterterms as  
the scale that defines the subtraction of ultraviolet divergences in loop amplitudes by expressing them in terms of the physical scale dependent parameters of the theory. A typical choice
for $\mu_R, \mu_F$ should be a scale close to the hard scale $Q$ which characterises the parton-parton interaction, and, as the scale is varied, the perturbative coefficients at higher order in~(\ref{eq:Xsec})
change in such way that the cross section to all orders is independent of the scale choice,
\begin{equation}
\frac{\partial\sigma}{\partial\mu_F}=\frac{\partial\sigma}{\partial\mu_R}=0\;.
\end{equation}
However, the scale variations of the cross section truncated at a given order only produces copies of the lower order terms and are not necessarily a reliable predictor of the size of the corrections at higher order.

In addition to reducing the sensitivity of the theory prediction on the unphysical scales, the inclusion of higher order corrections significantly improves the accuracy of 
the theory prediction for several reasons. At higher orders new partonic channels that do not contribute at leading order (LO) can appear.
Moreover, by including the effects due to emission of additional partons the phase space of the underlying Born process is enlarged and the parton-level prediction becomes more sensitive to the size of the jet resolution parameter and other details of the jet finding algorithm. 

The combination of all these effects can therefore lead to large radiative corrections that alter both the shape and normalisation of the prediction. In this talk
we will concentrate on the dijet process at hadron colliders and study the impact of the NNLO QCD corrections to the prediction in the gluons-only channel at full colour. 
 
\section{Dijet production}
Single jet inclusive jet and dijet observables are the most fundamental QCD processes measured at hadron colliders. They probe the basic parton-parton scattering in $2\to2$
kinematics, and thus allow for a determination of the parton distribution functions in the proton and for a direct probe of the strong coupling constant $\alpha_s$ up 
to the highest energy scales that can be attained in collider experiments. 

In the single jet inclusive cross section, each identified jet in an event contributes individually. 
The exclusive dijet cross section consists of all events with exactly two identified jets.
These cross sections have been studied as functions of different kinematic variables: the transverse momentum and rapidity of the jets (of any jet for the single jet 
inclusive distribution, or of the two largest transverse momentum jets for the dijet distributions). Precision measurements of single jet and dijet cross sections have been
performed by CDF~\cite{Aaltonen:2008eq} and D0~\cite{Abazov:2008ae} at the Tevatron and by ATLAS~\cite{Atlas} and 
CMS~\cite{CMS} at the LHC. 

The state of the art of the theoretical predictions for these observables are accurate to next-to-leading order (NLO) in QCD~\cite{NLO} (with the inclusion of shower
effects in~\cite{powheg2j}) and in the electroweak theory~\cite{Dittmaier:2012kx}. In this talk we
present results for the jet cross sections at NNLO accuracy in QCD in the gluons-only channel at full colour~\cite{nnlojet2}. 
To perform the calculation we have employed the antenna subtraction scheme~\cite{ourant} for the analytic cancellation of infra-red (IR) singularities at NNLO
extended to the case of processes with coloured particles in the initial state~\cite{ourant2}. As demonstrated in~\cite{nnlojet1}, using the antenna subtraction scheme the explicit
$\epsilon$-poles in the dimensional regularization parameter of one- and two-loop matrix elements entering this calculation are cancelled in analytic and local form
against the $\epsilon$-poles of the integrated antenna subtraction terms thereby enabling the computation of jet cross sections at hadrons colliders at NNLO accuracy. This allows the combination
of the two-loop virtual corrections to the basic $2\to2$ process~\cite{twol} together with the one-loop virtual corrections to the single real radiation $2\to3$ process~\cite{onelv} 
and the double real radiation $2\to4$ process at tree-level~\cite{real} in a parton-level event generator NNLOJET. Details of the calculation of the leading and sub-leading colour contributions
can be found in~\cite{nnlojet2}.

\section{Numerical results}
In this section we present numerical results for the partonic subprocess $gg\to gg+X$ at NNLO retaining full dependence on the number colours from proton-proton collisions at a 
center of mass energy of $\sqrt{s}=8$ TeV. Jets are identified using the anti-k$_{T}$ algorithm with resolution parameter R = 0.7. Jets are accepted at central rapidity |y| < 4.4, and ordered in transverse momentum.
An event is retained if the leading jet has $p_{T1} > 80$ GeV. Factorization and renormalization scales ($\mu_{F}$ and $\mu_{R}$) are chosen dynamically on an event-by-event basis.
As default value, we set $\mu_F$ = $\mu_R = \mu$ and set $\mu$ equal to the transverse momentum of the leading jet so that $\mu = p_{T1}$.

\begin{figure}[htp]
\centering
\includegraphics[width=0.71\textwidth]{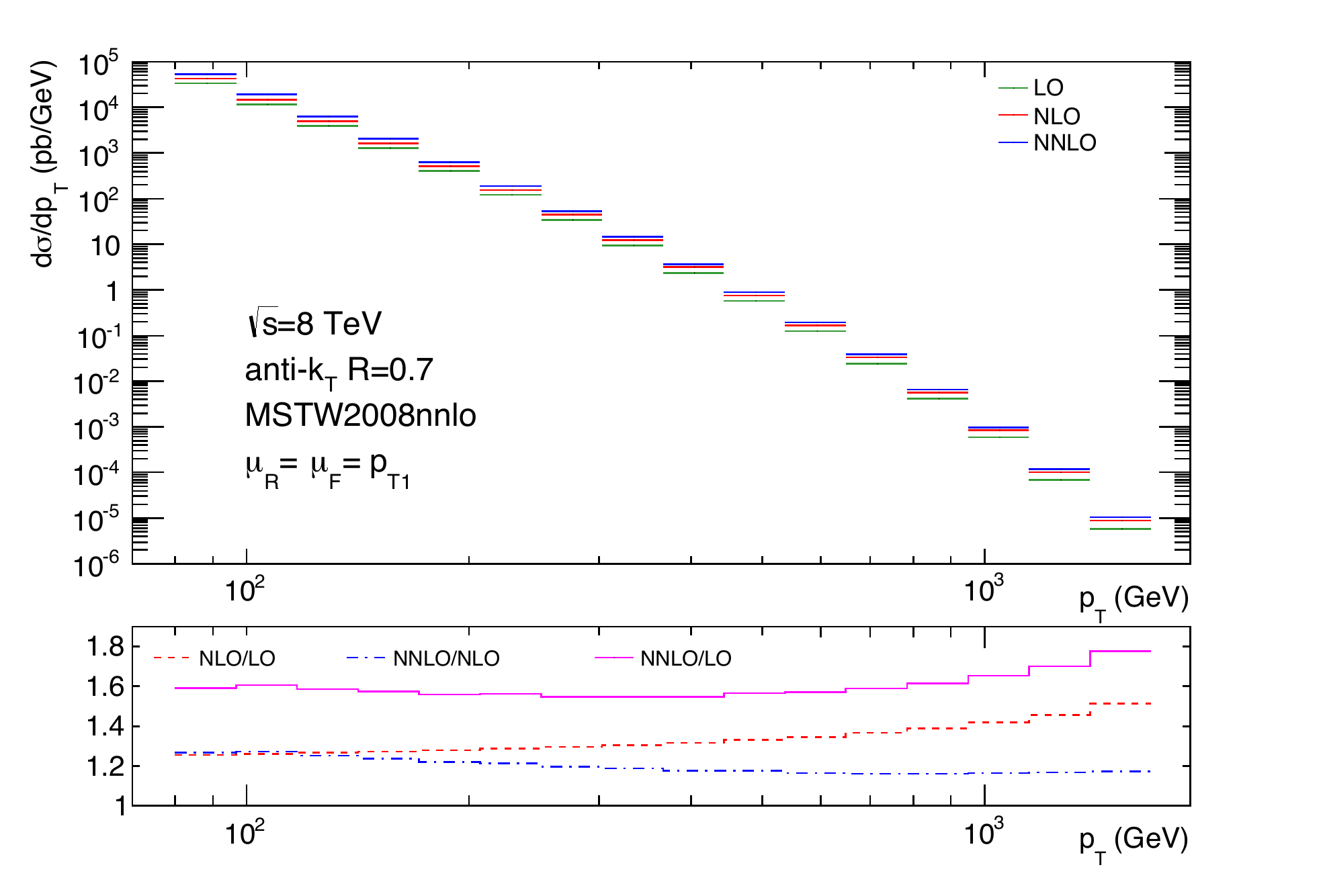}
\includegraphics[width=0.61\textwidth]{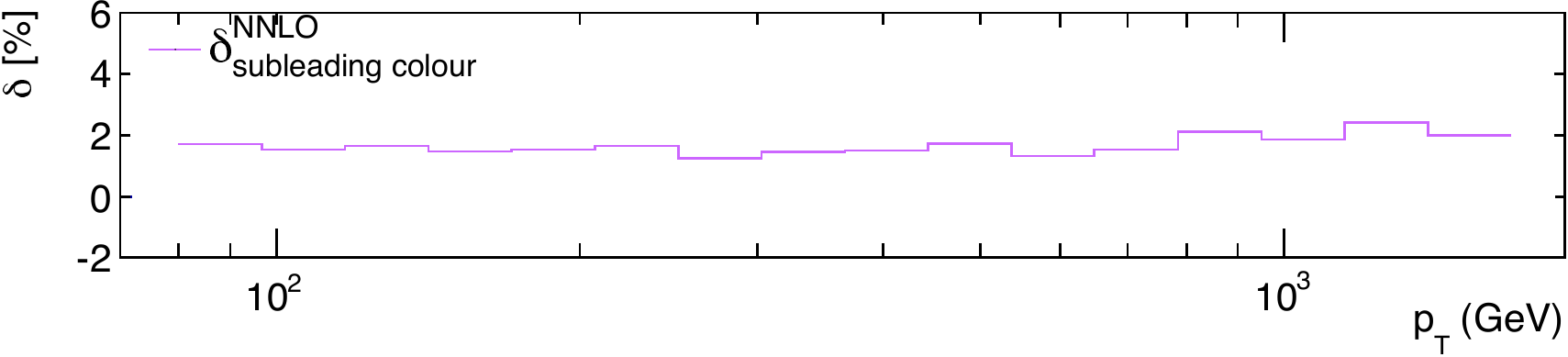}\hspace{0.7cm}
\caption{Inclusive jet $p_T$ distribution, $d\sigma/dp_T$, for jets constructed with the anti-$k_T$ algorithm with $R=0.7$ and with $p_T > 80$~GeV, $|y| < 4.4$ and $\sqrt{s} = 8$~TeV at NNLO (blue), NLO (red) and LO (dark-green).
The middle panel shows the ratios of NNLO, NLO and LO cross sections. The lower panel shows the subleading colour contribution to the full colour NNLO cross section.}
\label{fig:dsdet}
\end{figure}

In Fig.~\ref{fig:dsdet} we present the single jet inclusive cross section as a function of the jet $p_{T}$ at each order in perturbation theory. Respectively in dark-green, red and blue we show the LO, NLO and NNLO 
cross sections evaluated with the MSTW2008NNLO~\cite{mstw} gluon distribution function.
The middle panel shows ratios of NNLO, NLO and LO cross sections and we can observe that the NNLO corrections increase the NLO cross section between 16-26\% and stabilise the NLO/LO k-factor growth with $p_{T}$.
Finally in the lower panel we present the contribution of the subleading colour piece to the full colour NNLO cross section to test the validity of the leading colour approximation.
The subleading colour contribution in this channel appears for the first time at NNLO and contributes with a flat 2\% increase of the leading colour NNLO cross section.

\begin{figure}[htb]
\centering
\includegraphics[width=0.71\textwidth]{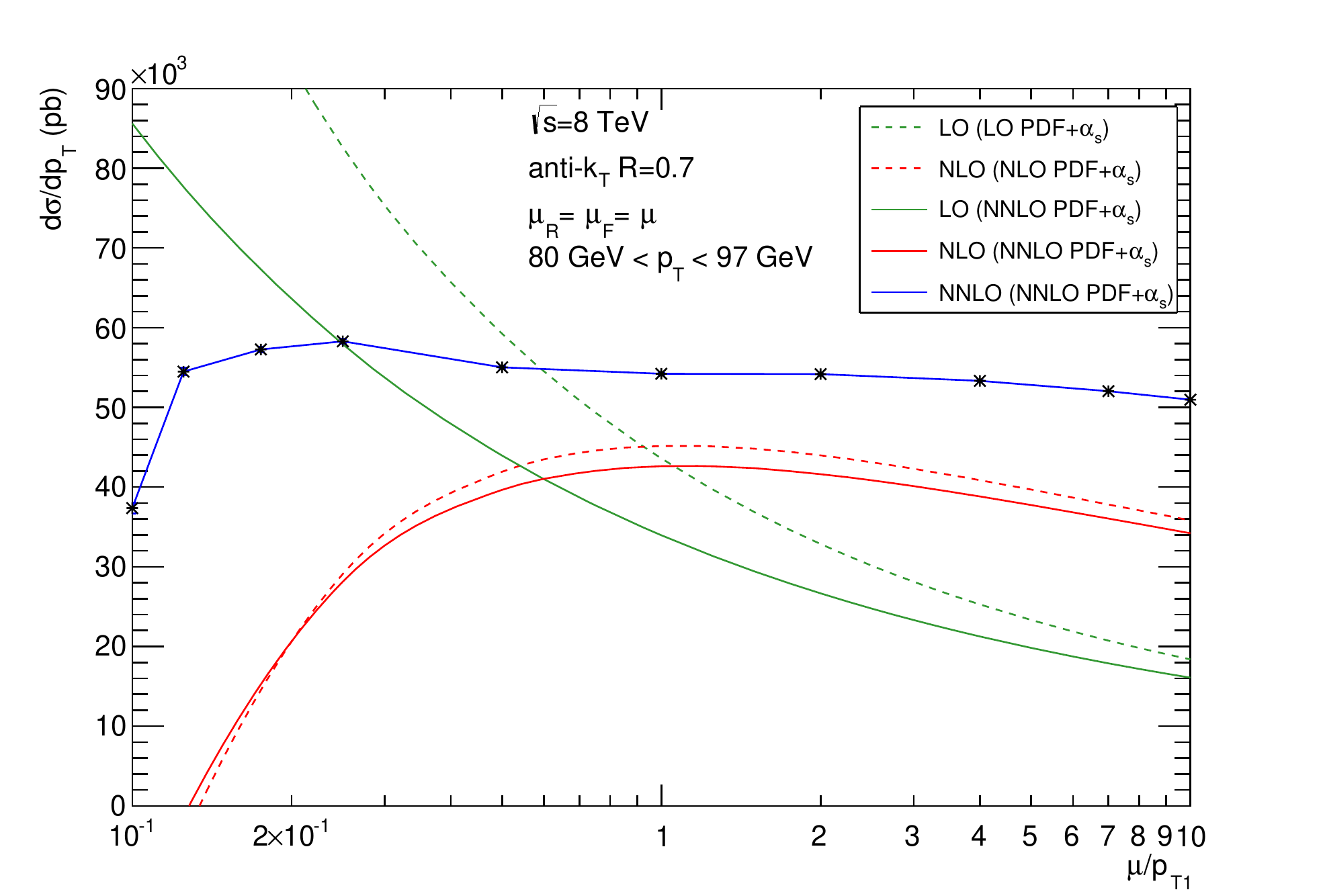}
\caption{Scale dependence of the NNLO (blue, solid), NLO (red, solid) and LO (green,solid) inclusive jet cross section for $pp$ collisions at $\sqrt{s}$=8 TeV in the gluons-only channel at full colour
for the anti-$k_{T}$ jet algorithm with R=0.7 and with $|y|<4.4$ and 80 GeV < $p_{T}$ < 97 GeV and evaluated with the MSTW08NNLO gluon distribution function. 
The LO (green, dashed) and NLO (red, dashed) contributions are evaluated with the corresponding LO and NLO sets.}
  \label{fig:scaledep}
\end{figure}
   
In Fig.~\ref{fig:scaledep} we study the single jet inclusive cross section at each order in perturbation theory as a function of the renormalization and factorization scales. In the gluons-only channel
we set the number of active flavours to zero ($N_F$=0) in the calculation except in the PDF evolution and running of $\alpha_s$ which are provided by LHAPDF~\cite{Whalley:2005nh}. For this study we consider jets 
with 80 GeV < $p_{T}$ < 97 GeV integrated over rapidity and perform a scale variation of a factor of ten around the central scale choice, $\mu_R = \mu_F = p_{T1}$. 
We observe that the inclusion of higher order corrections significantly reduces the scale dependence and at NNLO we obtain 
a flat scale dependence with a residual theoretical uncertainty from scale variations of a few percent. In the same figure we show the NLO and LO predictions using
the MSTW08NNLO gluon distribution function which allows us to quantify the size of the genuine higher order contributions to the parton-level subprocess.  For comparison, we also show the 
LO and NLO predictions using the corresponding LO and NLO PDF sets. There is a large increase of the cross section at LO due to the fact that the
LO PDF comes with a larger value for $\alpha_s$.  A similar, but much less pronounced, effect is also observed at NLO.  We observe that, for our central scale choice, the  NNLO cross section cannot be predicted simply by scale variations at NLO.

\begin{figure}[h!]
\centering
\begin{minipage}[b]{0.45\linewidth}
  \includegraphics[width=0.9\textwidth]{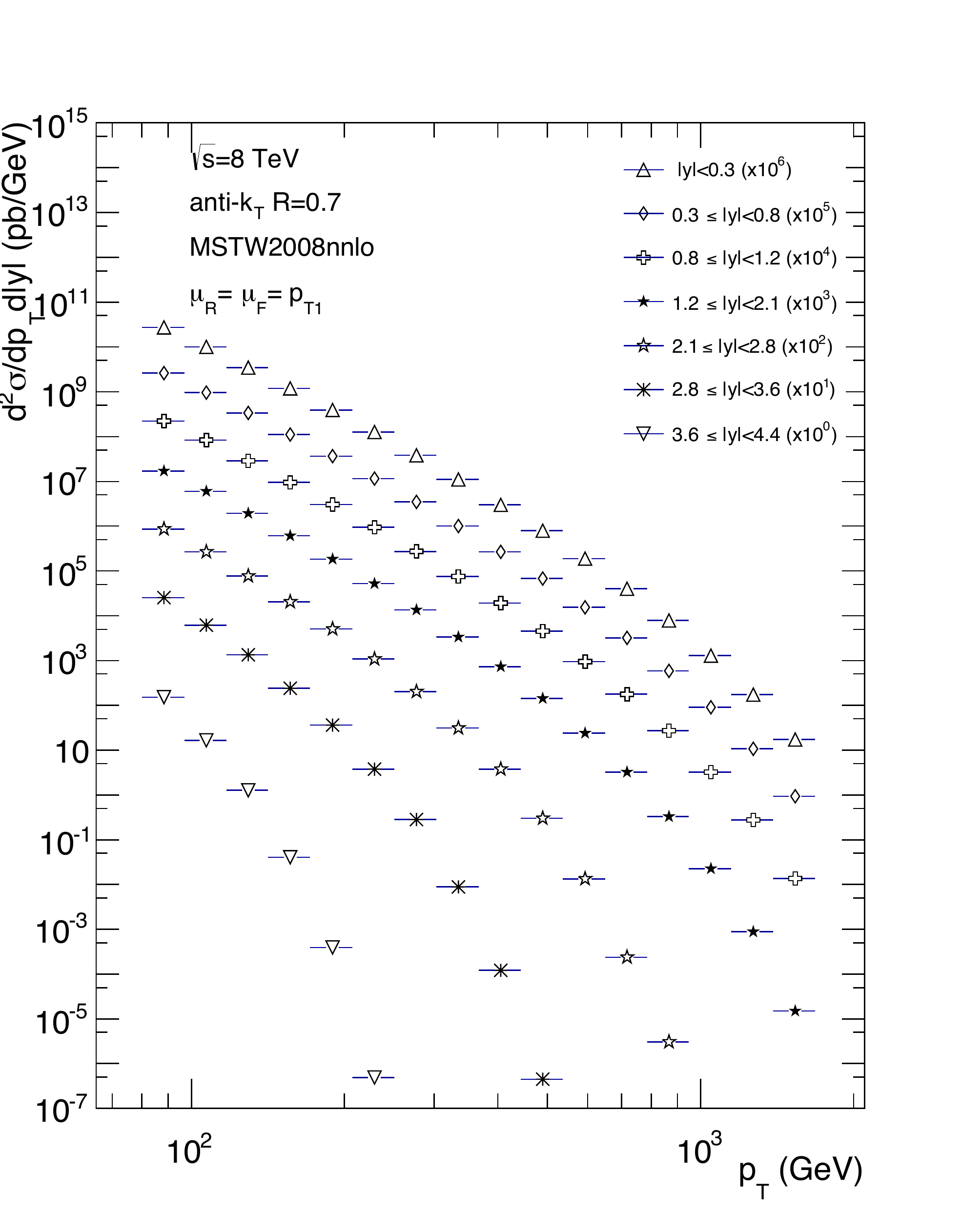}
\end{minipage}
\quad
\begin{minipage}[b]{0.45\linewidth}
  \includegraphics[width=1.1\textwidth]{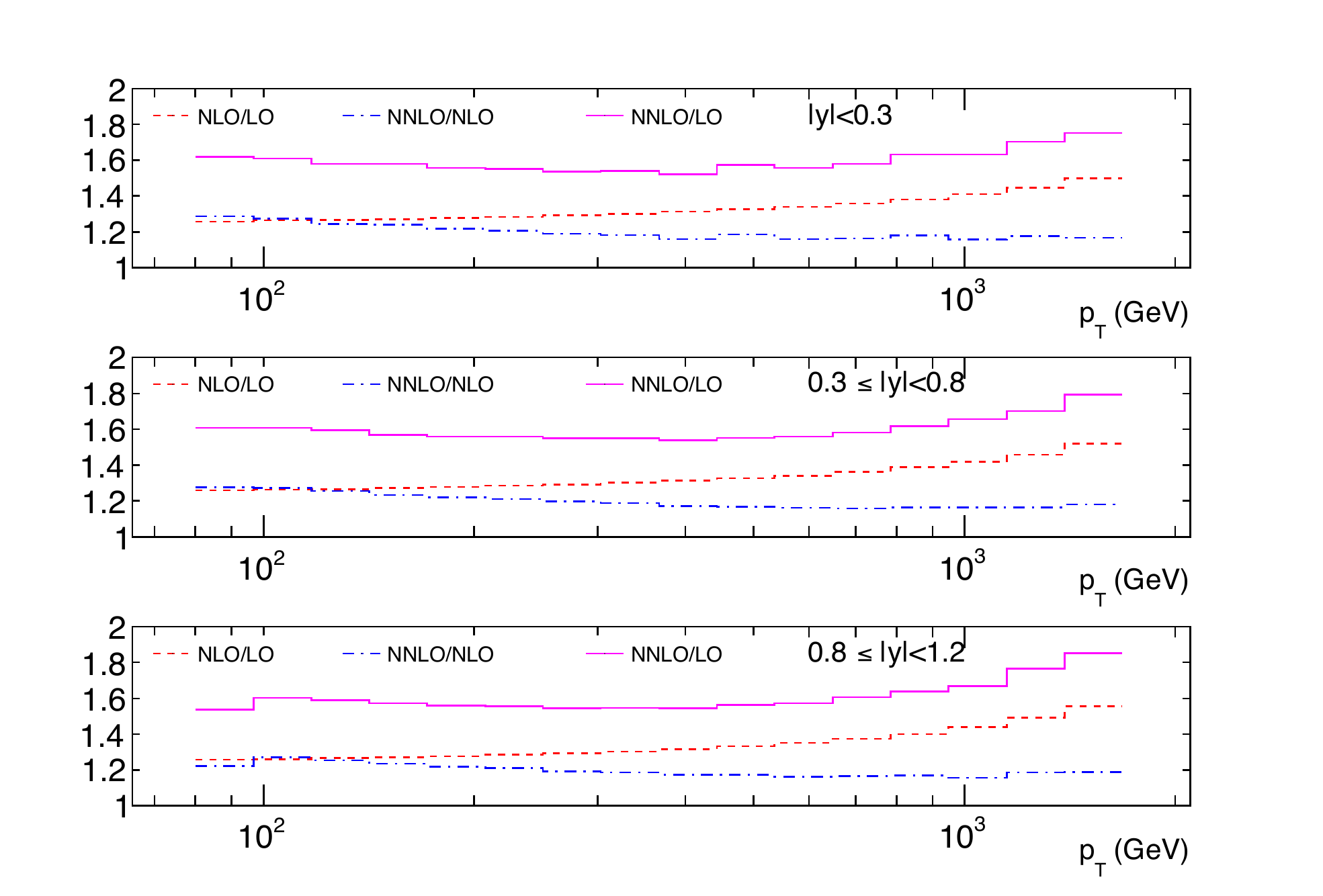}\\
\end{minipage}
  \caption{The left panel shows the doubly differential inclusive jet transverse energy distribution, $d^2\sigma/dp_T d|y|$, at $\sqrt{s} = 8$~TeV for the anti-$k_T$ algorithm with $R=0.7$ and for $p_T > 80$~GeV and various $|y|$ slices at NNLO. 
  The right panel shows the ratios of NNLO, NLO and LO cross sections for three rapidity slices: $|y | < 0.3$, $0.3 < |y| < 0.8$ and $0.8 < |y| < 1.2$.}
  \label{fig:d2sdetslice}
\end{figure}

In Fig.~\ref{fig:d2sdetslice} we present the inclusive jet cross section in double differential form at NNLO as it is measured at the LHC and Tevatron. The inclusive jet cross section
is computed in jet $p_{T}$ and rapidity bins over the range 0.0-4.4 covering central and forward jets. To quantify the impact of the NNLO correction we present the double differential $k$-factors containing 
ratios of NNLO, NLO and LO cross sections in the same figure. We observe that the NNLO correction increases the cross section between 26\% at low $p_{T}$ to 14\% at high $p_{T}$ with respect to the 
NLO calculation. This behaviour is similar for each of the three rapidity slices presented.

\begin{figure}[h!]
\centering
\begin{minipage}[b]{0.45\linewidth}
  \includegraphics[width=1.1\textwidth]{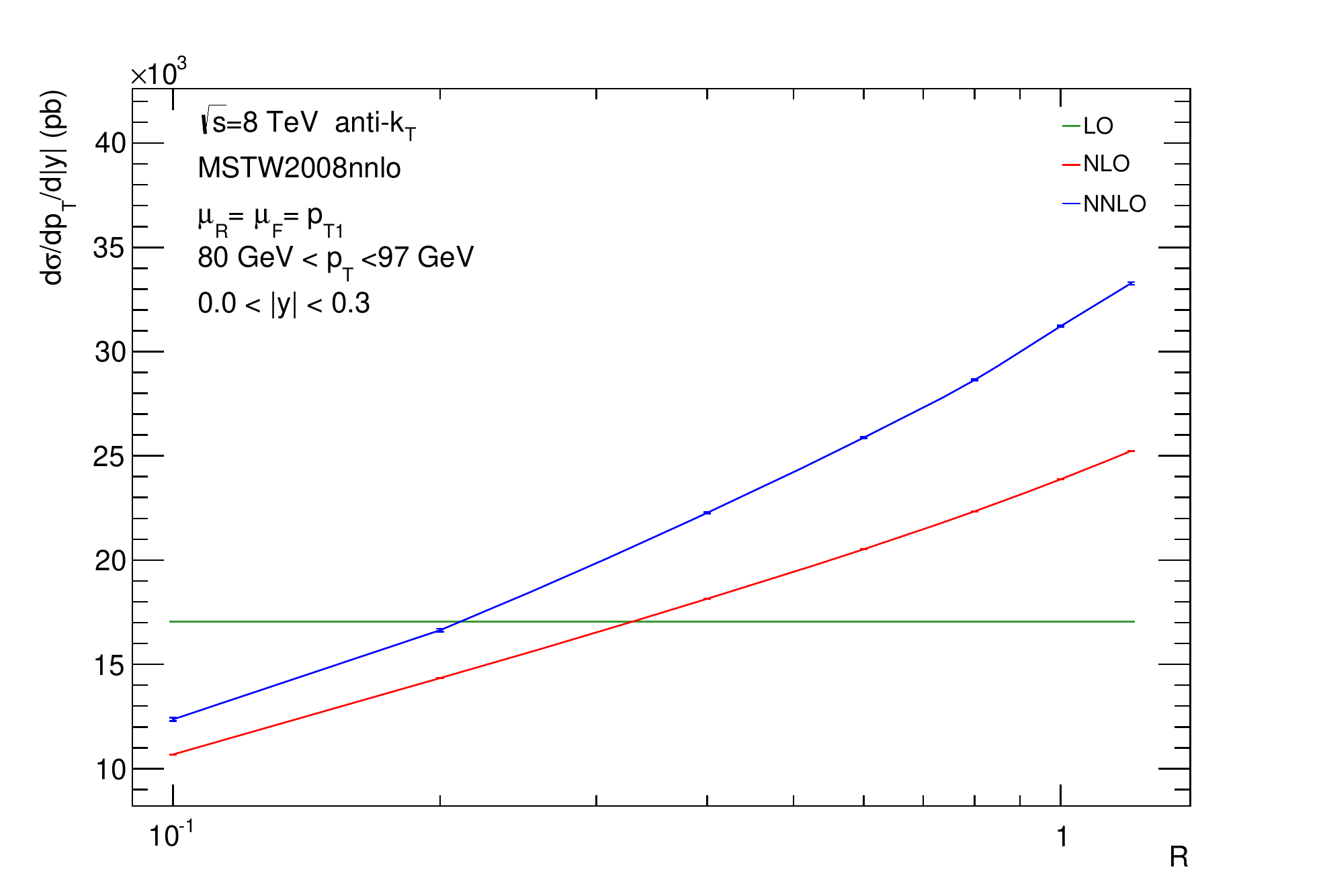}
\end{minipage}
\quad
\begin{minipage}[b]{0.45\linewidth}
  \includegraphics[width=1.1\textwidth]{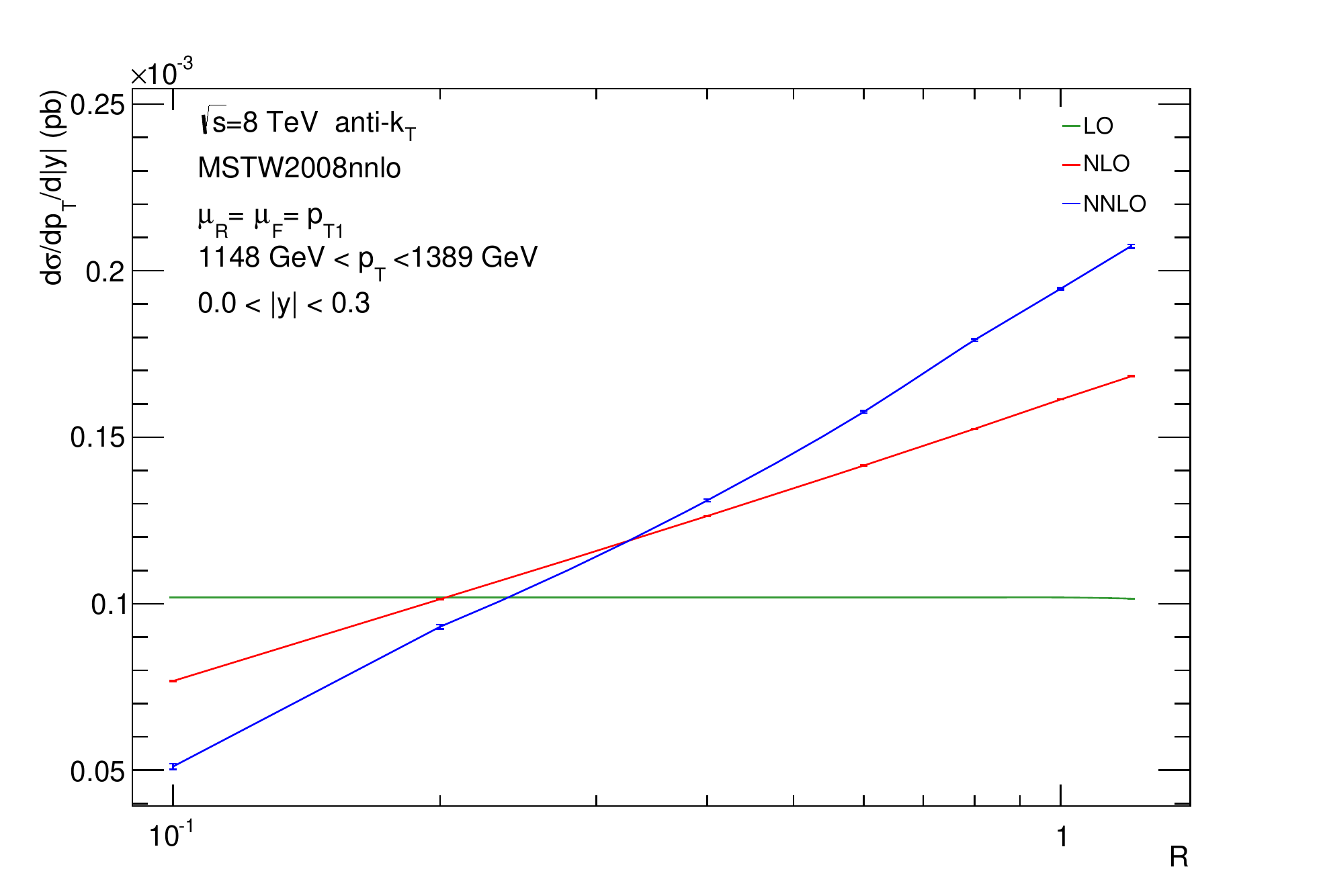}
\end{minipage}
  \caption{Single jet inclusive cross section versus resolution parameter $R$ of the anti-$k_T$ jet algorithm at $\sqrt{s}=8$~TeV, $80$~GeV $< p_T < 97$~GeV (left) and  $1148$~GeV $< p_T < 1389$~GeV (right) for
  $|y | < 0.3$ at NNLO (blue), NLO (red) and LO (green).}
  \label{fig:Rdep}
\end{figure}

As described in the introduction due to additional final-state radiation at the second order in perturbation theory we start to reconstruct the parton shower within the jet using exact matrix elements. 
In particular at this order up to three partons can form a single jet, leading to a better matching of the jet algorithm between theory and experiment. For this reason we studied, at each order
in perturbation theory, the single jet inclusive cross section as a function of the jet resolution parameter $R$ of the anti-$k_{T}$ jet algorithm and display the results in Fig.~\ref{fig:Rdep}. 
At LO each jet is modelled by one parton and the cross section is independent of the value of $R$. However at NLO for the first time additional parton radiation can be inside or outside the jet
and we can observe that the jet cross section increases with $R$. The NLO and NNLO predictions therefore increase with $R$ as more partons can be clustered to form the jet. For a given value of $R$, the relative size of the NLO and NNLO corrections depends on the transverse energy of the jet.

\begin{figure}[h!]
\centering
\begin{minipage}[b]{0.45\linewidth}
  \includegraphics[width=1.1\textwidth]{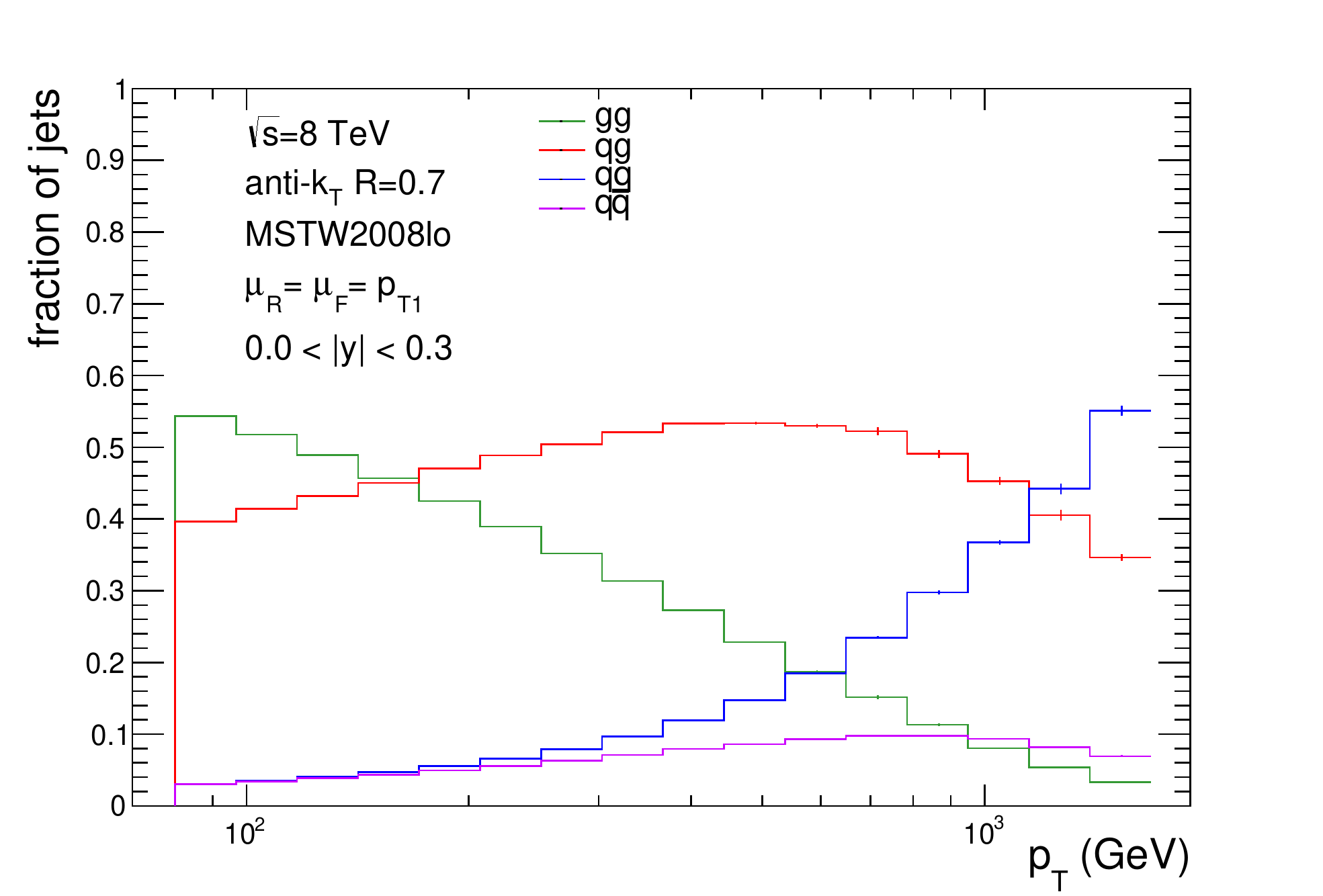}
\end{minipage}
\quad
\begin{minipage}[b]{0.45\linewidth}
  \includegraphics[width=1.1\textwidth]{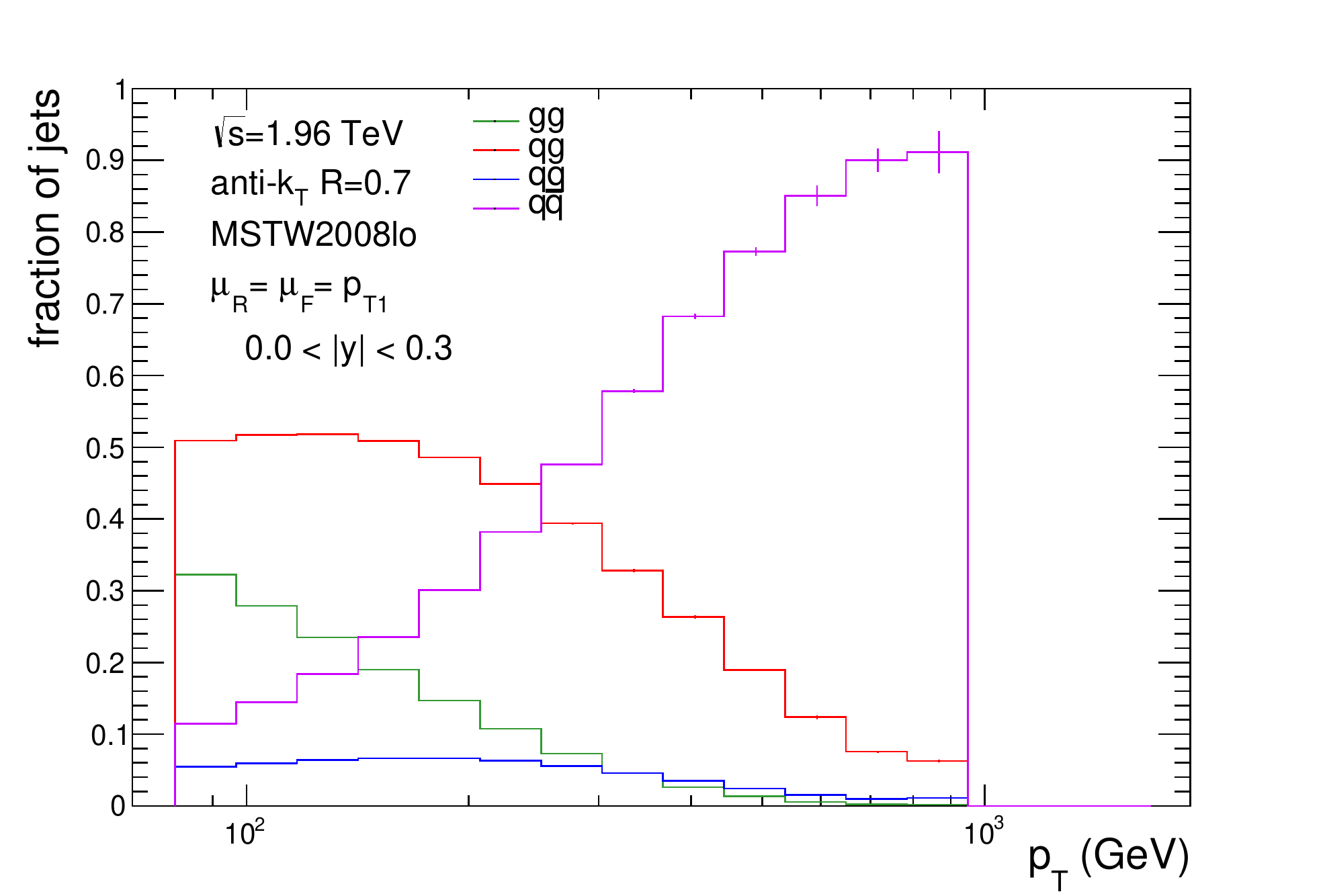}
\end{minipage}
  \caption{Fraction of jets per initial state scattering at the LHC (left) and at the Tevatron (right). The green, red, blue and magenta curves correspond to gluon-gluon scattering, quark-gluon scattering, quark-quark and
  quark-antiquark scattering respectively.}
  \label{fig:jetfrac}
\end{figure}

To conclude, we note that jets in hadronic collisions can be produced through a variety of different partonic subprocesses, of which the all-gluon process is but one. For this reason
it is useful to provide an estimate of how much of the full hadronic jet production proceeds through the gluons-only channel. In Fig.~\ref{fig:jetfrac} we show the fraction of jets per initial state
scattering at LO summing over final states at the LHC (left) and the Tevatron (right) for centrally produced jets $|y|<0.3$ using the MSTW2008LO PDF set. The $gg$ channel dominates at the LHC at low $p_{T}$ whereas at high 
$p_{T}$ the dominant processes are $qq$ and $qg$ scattering. The $qg$ channel has a contribution between 40-50\% across the whole $p_{T}$ range making it the second
most dominant channel at the LHC. This is not the case at the Tevatron where $qg$ scattering is the dominant channel at low and moderate $p_T$ 
and the high-$p_{T}$ jet production is completely dominated by $q\bar{q}$ scattering.

\section{Conclusions}
In this talk we have presented new results in the purely gluonic channel for the calculation of the fully differential dijet production process at hadron colliders at NNLO. 
This partonic subprocess is the most challenging one from the theoretical point of view since the gluons-only channel contains the most intricate IR structure of all partonic processes that contribute
to the production of jets. Moreover, results for this channel are now available in full colour demonstrating the capability of the antenna subtraction method to deal with colour-correlated matrix elements at NNLO.
For these reasons we believe we can tackle successfully the remaining partonic processes involving two quarks, four quarks and six-quarks and look forward to comparing our results directly with experimental data. 

\acknowledgments{
This research was supported by the Swiss National Science Foundation
(SNF) under contract PP00P2-139192, in part by the European Commission through the
`LHCPhenoNet' Initial Training Network PITN-GA-2010-264564 and in part by the UK Science
and Technology Facilities Council through grant ST/G000905/1.}

\bibliography{ref}

\end{document}